\documentclass[preprint]{elsarticle}

\usepackage{graphicx,lineno,hyperref,amsmath,amsfonts,amssymb,algorithm,algorithmicx,
booktabs,algcompatible,subcaption,epsfig}
\modulolinenumbers[250]

\algdef{SE}[DOWHILE]{Do}{doWhile}{\algorithmicdo}[1]{\algorithmicwhile\ #1}%

\journal{a journal}

\begin{document}

\begin{frontmatter}

\title{On the visualization of the detected communities in dynamic networks: A case study of Twitter's network}


\author[mymainaddress]{Youcef Abdelsadek\corref{mycorrespondingauthor}}
\cortext[mycorrespondingauthor]{Corresponding author}
\ead{youcef.abdelsadek@univ-lorraine.fr}

\author[mymainaddress]{Kamel Chelghoum}
\ead{kamel.chelghoum@univ-lorraine.fr}

\author[mymainaddress]{Francine Herrmann}
\ead{francine.herrmann@univ-lorraine.fr}

\author[mymainaddress]{Imed Kacem}
\ead{imed.kacem@univ-lorraine.fr}

\author[mysecondaryaddress]{Beno\^it Otjacques}
\ead{benoit.otjacques@list.lu}

\address[mymainaddress]{Laboratoire de Conception, Optimisation et Mod\'elisation des Syst\`emes\\ LCOMS EA 7306, Universit\'e de Lorraine, Metz, France}
\address[mysecondaryaddress]{e-Science Research Unit, Environmental Research and Innovation\\ Luxembourg Institute of Science and Technology, Belvaux, Luxembourg}

\begin{abstract}
Understanding the information behind social relationships represented by a network is very challenging, especially, when the social interactions change over time inducing updates on the network topology. In this context, this paper proposes an approach for analysing dynamic social networks, more precisely for Twitter's network. Our approach relies on two complementary steps: (i) an online community identification based on a dynamic community detection algorithm called \textit{Dyci}. The main idea of \textit{Dyci} is to track whether a connected component of the weighted graph becomes weak over time, in order to merge it with the "dominant" neighbour community. Additionally, (ii) a community visualization is provided by our visualization tool called \textit{NLCOMS}, which combines between two methods of dynamic network visualization. In order to assess the efficiency and the applicability of the proposed approach, we consider real-world data of the ANR-Info-RSN project, which deals with community analysis in Twitter. 
\end{abstract}


\end{frontmatter}

\linenumbers

\section{Introduction}

With the popularization of social networks like Twitter, an exponential quantity of data is generated and has led to the arise of relatively new field called, social network analysis \cite{wasserman1994}. These data are increasing each day, and the existing approaches which are not considering the dynamic nature of data would suffer from the scalability issue. These social relationships can be appropriately represented by networks \cite[Chapter~10]{LeskovecRU14}. For example, e can cite social relationships like the friendship, the follow in social blogs or the information sharing in social media. Commonly, social relationships are not static but could change over time with, appearing and/or disappearing relationships for binary relationships (e.g., friendship) or increasing and/or decreasing for weighted relationships (e.g., the number of times where two persons share information). This evolving complex networks are called dynamic graphs. In this paper, we consider edge weighted dynamic graphs, where a weight is assigned to each edge of the graph to model a well-known relation in Twitter, which is the retweet. The latter occurs when a Twitter user republishes an original tweet of another Twitter user. Therefore, the edge weight corresponds to the number of times where a retweet is observed between two Twitter users. One among the objectives of this paper is to reveal the more interconnected group of nodes, known as communities, over time. In other words, the purpose of this work is to figure out the evolution of the most active groups of Twitter users. 

\begin{figure}[!t]
\begin{center}
\includegraphics[scale=0.27]{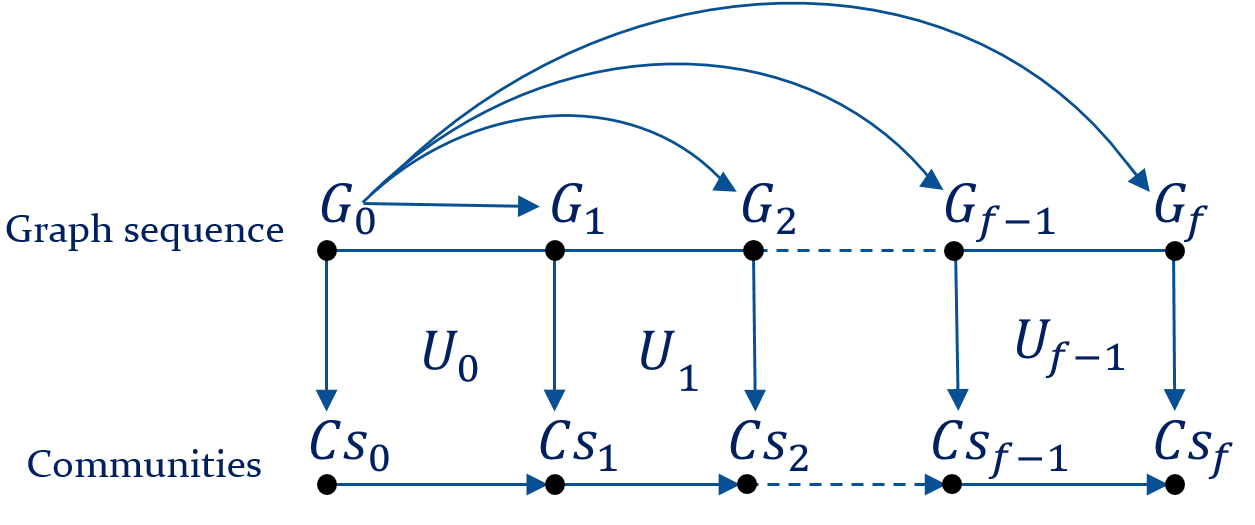}
\end{center}
\caption{The graph sequence of a dynamic graph}%
\label{dynChang}%
\end{figure}

More formally, a dynamic graph of an initial graph $ \mathit{G_{0}} $ can be seen as a sequence of static graphs \cite{Diehl2002}, denoted by $\mathit{G_{s}} = (\mathit{G_{0}}, \mathit{G_{1},\ldots, \mathit{G_{f}}})$ with $ f $ snapshots as results of the updates $\mathit{U_{s}} = (\mathit{U_{0}}, \mathit{U_{1},\ldots,} \mathit{U_{f-1}})$ as illustrated in Figure \ref{dynChang}. We denote by $\mathit{N_{t}}$, $\mathit{E_{t}}$, $\mathit{E_{t}^{w}}$, $\mathit{N_{s}}$ and $\mathit{E_{s}}$ respectively, the set of nodes of size $ v $ at instant \textit{t}, the set of edges at instant \textit{t}, the set of edge weights at instant \textit{t} of $ \mathit{G_{t}} $, the set of nodes of the whole $\mathit{G_{s}}$ and the set of edges of the whole $\mathit{G_{s}}$. Furthermore, the set of updates $ \mathit{U_{t}} $ varies in terms of the impact they cause to the current network structure. As an instance, the impact of adding a new node and those of updating the weight of an existing edge differ. We point out that weights can be assigned to the nodes also, with node weight update scenario for the dynamic context, which is not considered in this paper. The following updates cases describes the repercussion on $\mathit{N_{t}}$, $\mathit{E_{t}}$ and $\mathit{E_{t}^{w}}$ after each update scenario.

\begin{enumerate}
\item \textbf{Structural updates}:
\begin{enumerate} 
\item \textbf{Node removing}: An old node $ \mathit{on} $ is removed, $ \mathit{N}_{t+1} \leftarrow \mathit{N}_{t} \setminus \left\lbrace \mathit{on} \right\rbrace $ with the related edges.
\item \textbf{Edge removing}: An old edge $ \mathit{oe} $ is removed, $ \mathit{E}_{t+1} \leftarrow \mathit{E}_{t} \setminus \left\lbrace \mathit{oe} \right\rbrace $.
\item \textbf{Node addition}: A new node $ \mathit{nv} $ is added, $ \mathit{N}_{t+1} \leftarrow \left\lbrace \mathit{nv} \right\rbrace  \cup \mathit{N}_{t}$ with the related edges.
\item \textbf{Edge addition}: A new edge $ \mathit{ne} $ is added, $ \mathit{E}_{t+1} \leftarrow  \left\lbrace \mathit{ne} \right\rbrace \cup \mathit{E}_{t}$.
\end{enumerate}
\item \textbf{Attributes updates}: 
\begin{enumerate} 
\item \textbf{Edge weight updating}: An new edge weight $ \mathit{ne}^{w} $ of an old edge weight $ \mathit{oe}^{w} $ is updated, $ \mathit{E}_{t+1}^{w}  \leftarrow \left( \mathit{E}_{t}^{w} \setminus \left\lbrace \mathit{oe}^{w} \right\rbrace \right) \cup \left\lbrace \mathit{ne}^{w} \right\rbrace $.
\end{enumerate}
\end{enumerate}

Furthermore, one among the important objectives in social network analysis is to reveal the semantic aspects behind the network topology. This objective can be reached by identifying the highly intra-connected groups of nodes or communities, which are in the opposite poorly inter-connected, known as community detection problem \cite{Fortunato201075}. This problem is very challenging and can be found in several domains. For example, it can be met in sociology \cite{freeman04} where we aim to study the common characteristics of social groups and what makes groups of people together in a community. However, the community detection problem becomes trickier when the network topology changes over time. The changes that might occur to the community set $\mathit{Cs_{s}} = (\mathit{Cs_{0}}, \mathit{Cs_{1},\ldots, \mathit{Cs_{f}}})$ can affect either the structure, the attributes (i.e., weights) or also both of them \cite{Harary1997}. Consequently, how to analyse the evolution of the communities structure over time? To answer this question, one need to devise an algorithm, which relies on the graph features and which takes advantage from the previously identified communities by avoiding the community identification from scratch at each instant. As a concrete example, an analyst needs to understand how the information is shared in Twitter by understanding the role of each Twitter user within its community and outside of its community. To fulfil this need, one have to detect the communities of the analyst's time point of interest and to follow the community's member evolution with new members joining/leaving the studied communities. In this context, a trade-off between efficiency and response time is necessary to detect the community evolution over time. 

As previously mentioned, one among the major task that we would like to carry out is to understand the community structure behaviour over time. This drives us to deal with very challenging field, which is related to the representation of relational and dynamic data using node-link diagrams as depictions, called dynamic graph drawing \cite{Branke}. Recently, dynamic graph drawing and visualization gain more interest from the scientific community. They raises several new challenges additionally to challenges of the static context leading to harder tasks. For example, how to recognize with least cognitive effort the changes that occur in the graph structure? If only a small part of the graph changes, how to conserve the stability of the other parts of the graph? In order to fulfil these requirements, one have to make a trade-off between stability and efficiency. This is what we try to target in order to reach the aforementioned objectives. In the context of the previously introduced Twitter's network case. The Twitter members can retweet each others at different points of time and with different frequencies. Indeed, the analyst could want to perceive user's activity with the induced consequences on the community partition. 

The remaining part of this paper is organized as follows: In Section \ref{relatedWork}, the related work of the addressed topics are presented. Section \ref{Dyci} introduces the \textit{Dyci} algorithm for community identification in dynamic graphs while a genetic algorithm for the community detection problem is introduced in Section \ref{GA}. Section \ref{experiments} presents the case study and the conducted experiments on real-world data of the ANR Info-RSN project. In Section \ref{vis}, the description of the used methods for dynamic graph visualization is presented. Finally, Section \ref{conlusion} concludes this paper and discuss the future improvements of this work.

\section{Related work}
\label{relatedWork}

\subsection{Community detection}

This section presents some related works for the dynamic community detection. There are more algorithms for the static version of this problem in the literature compared to the dynamic case, especially, for those considering weighted edges. The static community detection algorithms can be divided into two families of algorithms: the divisive family (top-down) \cite{NewGir04} and the agglomerative family (bottom-up) \cite{blondel2008}. The dynamic community detection problem was proved in \cite{1281269} to be \textit{NP}-complete and \textit{APX}-hard. For the unweighed dynamic community detection, the authors of in \cite{Takaffoli2011} propose a matching algorithm to detect similar communities over the snapshots of the graph sequence, forming a meta-community, which is the sequence of these identical communities. Agglomerative modularity-based approach are considered in \cite{BansalBP10}, \cite{AktuncTOD15} and \cite{NguyenDXT11}. The authors of \cite{NguyenDXT11} use a physical principle with forces, which retains a node to stay in its community against attracting forces of the other communities. Furthermore, game-theory analogy is used in \cite{AlvariHS14}. In the latter, each node of the graph is considered as an agent, which maximizes its utility function. A set of predefined agent actions is set initially. The system ends when all agents choose their best community belonging (i.e., which maximizes the utility function). Relying on the colouring problem, a constant-approximation algorithm was proposed in \cite{TantipathananandhB09}. The authors of \cite{HopKhaKulSel04} deal with changes tracking of communities in large networks. They propose an approach which uses agglomerative clustering to examine the evolution of the community structure over time by identifying stable communities after several cluster running. In \cite{Greene2010} a model is described which tracks communities over time, those are characterised by a set of events. Regarding the weighted version of this problem, label propagation is used \cite{abs-1305-2006}. The idea of this algorithm is to allow a specific node to change its community label taking into account its adjacent nodes labels. 

\subsection{Drawing and visualization of dynamic graphs}

This section reports some of the existing dynamic graphs drawing \cite{Branke} and visualisation techniques \cite{beck2014}. Force-directed algorithms are well-known to be appropriate for static graph drawing. They produce pleasing layouts which meet the graph drawing aesthetic criteria. For example, the edge crossing minimization and the uniformity of edge length have been considered. Furthermore, they are also appropriate for drawing dynamic graphs leading to additional aesthetic criteria \cite{beck2009}. One among the objectives of drawing dynamic graphs is to maintain the familiarized graph structure that the user built up over time, so-called user mental map \cite{ArchambaultP13}. Generally, its preservation through the successive snapshots helps the user to stay familiar with the graph structure. The mental map preservation also avoids to get the user away from its original task, which is the graph structure analysis. In \cite{FrishmanT07}, the authors devised a GPU-based implementation algorithm for on-line dynamic drawing. The idea of their revisited force-directed algorithm is to assign weights to graph nodes by encoding the node's movement rigidity. This means that the greater the node weight is, the less the node is flexible. The latter ensures the user mental map preservation.

After the graph is embedded, comes the visualization issue. Indeed, visualizing dynamic graphs and dynamic data in general is challenging \cite{AignerMST11}. There are two major techniques to visualize dynamic graphs. The first one relies on the physical time differing the layout of each snapshot, whereas in the second technique, an axis is used to present the successive snapshots. Several experiments have been performed in order to asset the efficiency of each technique \cite{ArchambaultP13}\cite{ArchambaultPP11}. In the former technique, the whole depiction area is dedicated to one snapshot. Additionally, the animation can be used in order to provide smooth snapshots transition. For example, in \cite{ErtenHKWY03} the authors use a fade out effect for node removing. In the latter technique the depiction area is divided either in grid format or time-line, so-called small-multiple \cite{Tufte1990} where the user can follow the graph structure evolution by respecting a precedence constraint on an axis. Furthermore, these chronologically embedded snapshots can be visualized on a 2.5 dimensions with snapshot superposition \cite{PupyrevT10}. In \cite{ArchambaultPP10} the authors layout the union of two successive snapshots in one aggregated snapshot in order to highlight the difference between them relying on colors as visual variable, called difference map.

A combination of these techniques in multiple views is possible, like in Diffani \cite{RufiangeM13} with difference map and animation. 

\section{Dynamic community detection algorithm}
\label{Dyci}

\subsection{Notations and definitions}

Let us define $ c_{n_{i}}, IW, INW, WD $ and $ WI $ which, respectively, represent the community of the node $ n_{i} $, the intra-community weight, the inter-community weight, the weighted degree of a node and the weighted community-incidence of a node. These are presented in the following equations: 

\begin{equation}
IW_{c_{g}} = \sum_{n_{i} \in c_{g}} \sum_{n_{j} \in c_{g}} \frac{e^{w}_{n_{i}, n_{j}}}{2} 
\end{equation}
\begin{equation}
INW_{c_{g}, c_{h}} = \sum_{n_{i} \in c_{g}} \sum_{n_{j} \in c_{h}} e^{w}_{n_{i}, n_{j}}
\end{equation}
\begin{equation}
WD_{n_{i}} = \sum^{v}_{j=1} e^{w}_{n_{i}, n_{j}}
\end{equation}
\begin{equation}
WI(node, c_{g}) = \frac{\sum_{n_{i} \in c_{g}} e_{node, n_{i}}^{w}}{WD_{node}} 
\end{equation}

\subsection{\textit{Dyci} algorithm}

First of all, the static community detection algorithm proposed in \cite{abdelsadek} is applied on $ \mathit{G_{0}} $ as initial community partition $ \mathit{Cs_{0}} $ of \textit{Dyci}. The main idea of this algorithm consists in using a collection of triangles as a starting point. Then, adjacent communities are iteratively compared in terms of weights and merged when a merging condition holds. This iterative process ends when no community merging is observed. However, for the other snapshot of the graph sequence, \textit{Dyci} reacts depending on the update scenario as presented in the Algorithm \ref{DynamicComAlgo}. 

\begin{algorithm} [!h]
\caption{\textit{Dyci algorithm}}
\label{DynamicComAlgo}
\begin{algorithmic}
\REQUIRE $ \mathit{G_{t}} $, $ \mathit{Cs_{t}} $ and $ \mathit{U_{t}} $;
\ENSURE  $ \mathit{Cs_{t+1}} $.
\\ \textbf{BEGIN}
   \IF{$ \mathit{G_{t}} $ is $ \mathit{G_{0}} $} 
    	\STATE Launch the static community detection algorithm \textit{Tribase} \cite{abdelsadek}.
    \ELSE
     \FORALL {$ oldNode $ \textbf{in} $ \mathit{U}_{t}.nodeToRemove $}
		\STATE $ \mathit{Cs_{t}} \leftarrow $ \textit{NodeRemoving} $ (oldNode, \mathit{Cs_{t}}, \mathit{G_{t}})  $;
	 \ENDFOR
	 \FORALL {$ oldEdge $ \textbf{in} $ \mathit{U}_{t}.edgeToRemove $}
		\STATE $ \mathit{Cs_{t}} \leftarrow $ \textit{EdgeRemoving} $ (oldEdge, \mathit{Cs_{t}}, \mathit{G_{t}})  $;
	 \ENDFOR
	 \FORALL {$ newNode $ \textbf{in} $ \mathit{U}_{t}.nodeToAdd $}
		\STATE $ \mathit{Cs_{t}} \leftarrow $ \textit{NodeAddition} $ (newNode, \mathit{Cs_{t}}, \mathit{G_{t}})  $;
	 \ENDFOR	 
	 \FORALL {$ newEdge $ \textbf{in} $ \mathit{U}_{t}.edgeToAdd $}
		\STATE $ \mathit{Cs_{t}} \leftarrow $ \textit{EdgeAddition} $ (newEdge, \mathit{Cs_{t}}, \mathit{G_{t}})  $;	 
	 \ENDFOR	 
	 \FORALL {$ edgeWeightUpdate $ \textbf{in} $ \mathit{U}_{t}.edgeWeightUpdate $}	 	
	 	\STATE $ \mathit{Cs_{t}} \leftarrow $\textit{EdgeWeightUpdating} $ (newEdgeWeight, \mathit{Cs_{t}}, \mathit{G_{t}})  $;	
	 \ENDFOR	 
	\ENDIF
    \STATE \textbf{Return} $ \mathit{Cs_{t}} $;
   \\ \textbf{END.}
\end{algorithmic}
\end{algorithm}

The following subsections show how \textit{Dyci} reacts depending on the update scenario. Each update case is considered and presented in detail with the related illustrative example.

\subsubsection{ NodeRemoving (\textit{oldNode}):}
\begin{figure}[!h]
\centering
\begin{subfigure}{.5\textwidth}
  \centering
  \includegraphics[width=0.65\linewidth]{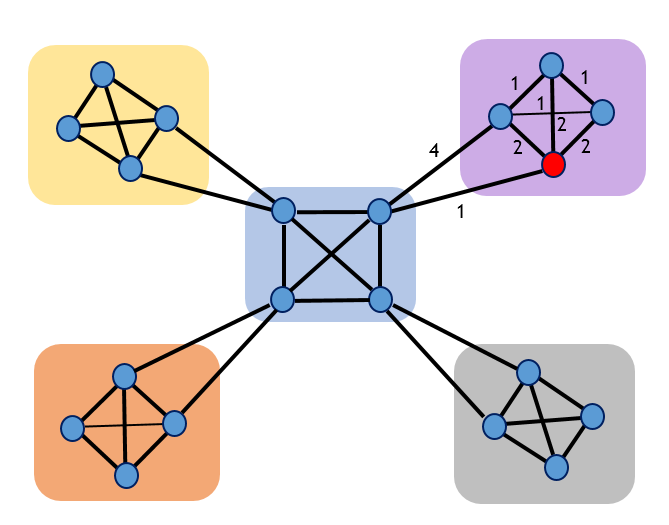}
  \caption{Before removing the red node}
  \label{fig:NRB}
\end{subfigure}%
\begin{subfigure}{.5\textwidth}
  \centering
  \includegraphics[width=0.65\linewidth]{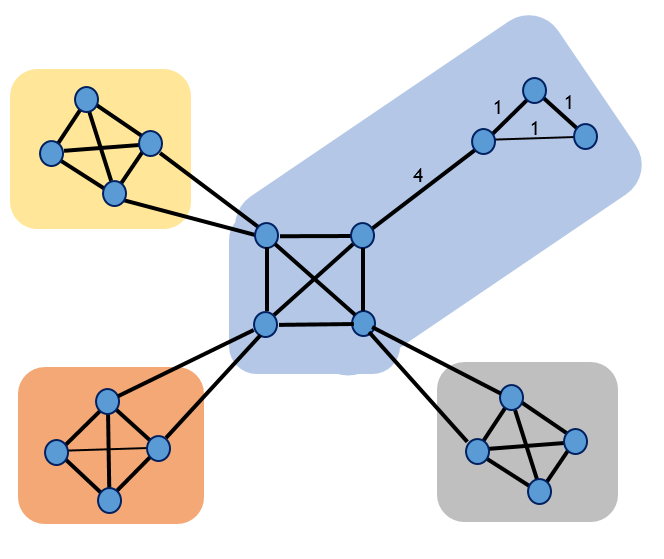}
  \caption{After removing the red node}
   \label{fig:NRA}
\end{subfigure}
\caption{Node removing scenario example}
\label{fig:NR}
\end{figure}

The main idea of the node removing case is to check whether the deletion of $ oldNode $ generates several connected components or reduces the $ IW(c_{oldNode}) $. To this end, \textit{Dyci} tests for each resulting connected component, noted $ CC $, whether it can form a community by it self or would be merged with an adjacent community, noted $ com $. In other words, \textit{Dyci} verifies whether Equation \ref{merginCondition1} holds or not. Figure \ref{fig:NR} gives an example of the node removing update scenario.

\begin{equation}
\label{merginCondition1}
INW_{com, CC} \geqslant IW_{CC}
\end{equation}

\subsubsection{ EdgeRemoving (\textit{oldEdge}):}
\begin{figure}[!b]
\centering
\begin{subfigure}{.5\textwidth}
  \centering
  \includegraphics[width=0.65\linewidth]{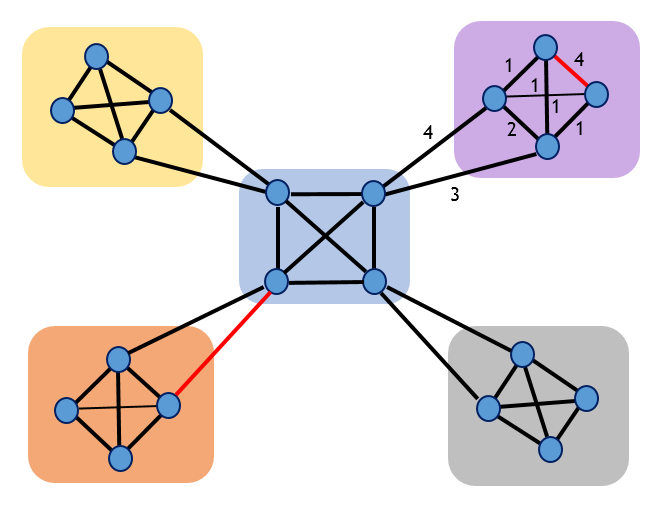}
  \caption{Before removing the red edge}
  \label{fig:ARB}
\end{subfigure}%
\begin{subfigure}{.5\textwidth}
  \centering
  \includegraphics[width=0.65\linewidth]{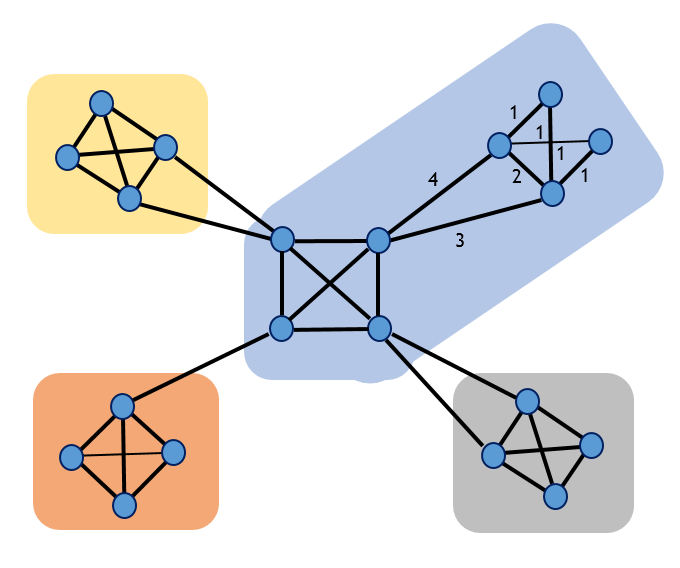}
  \caption{After removing the red edge}
  \label{fig:ARA}
\end{subfigure}
\caption{Edge removing scenario example}
\label{fig:AR}
\end{figure}

\textit{Dyci} handles the edge removing scenario as follows. Obviously, when an inter-community edge is removed, this reduces the inter-community weight leading to more community-like structure. However, the opposite case might lead to intra-community dividing in two connected components or a significant weight loss. To manage this second case, \textit{Dyci} compares the weights between each resulting connected component of \textit{oldEdge} deletion and their adjacent communities by Equation \ref{merginCondition1}. The edge removing update scenario is illustrated with an example in Figure \ref{fig:AR}. 

\subsubsection{ NodeAddition (\textit{newNode}):}

\begin{figure}[!h]
\centering
\begin{subfigure}{.5\textwidth}
  \centering
  \includegraphics[width=0.65\linewidth]{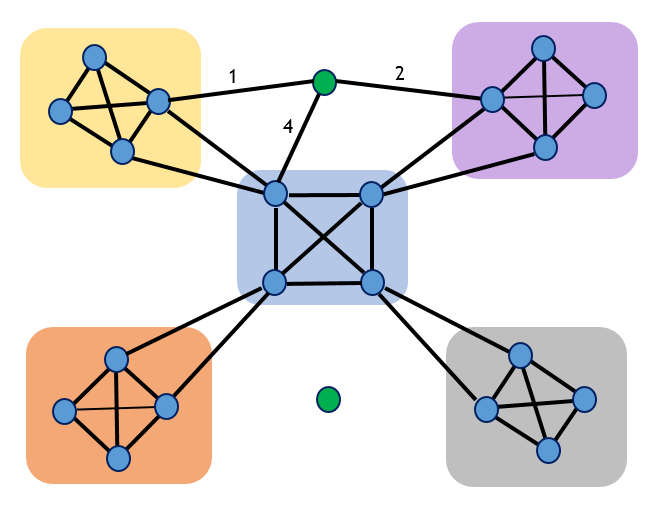}
  \caption{Before adding the green nodes}
  \label{fig:NAB}
\end{subfigure}%
\begin{subfigure}{.5\textwidth}
  \centering
  \includegraphics[width=0.65\linewidth]{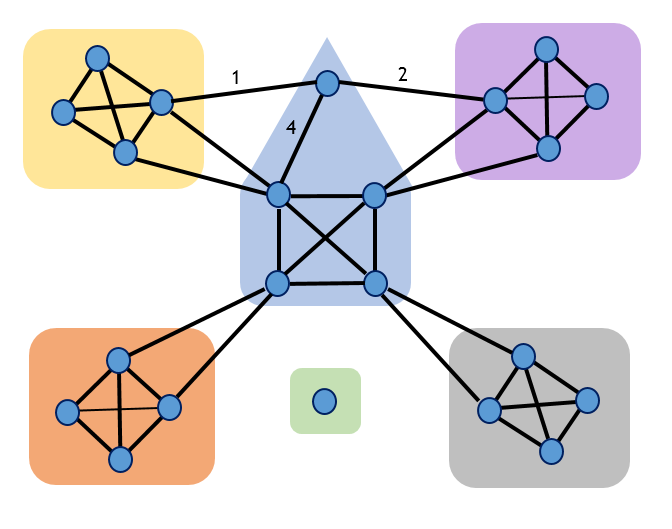}
  \caption{After adding the green nodes}
  \label{fig:NAA}
\end{subfigure}
\caption{Node addition scenario example}
\label{fig:NA}
\end{figure}

Two subcases can occur for node addition scenario. The first one, which can be viewed as trivial, is the subcase where $ newNode $ has no community edge incidence leading to an isolated community. In the second subcase, which is more common, $ newNode $ comes with many edges. For the latter, $ newNode $ is added to the community with the greatest $ WI(newNode, c)$, where $ c $ is a community adjacent to $ newNode $. Figure \ref{fig:NA} gives an example of the node addition update scenario.

\subsubsection{ EdgeAddition (\textit{newEdge}):}
\begin{figure}[!b]
\centering
\begin{subfigure}{.5\textwidth}
  \centering
  \includegraphics[width=0.65\linewidth]{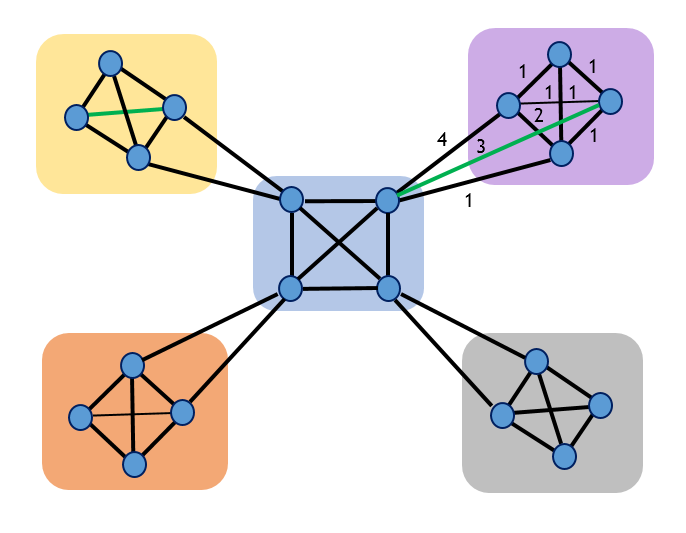}
  \caption{Before adding the green edges}
  \label{fig:AAB}
\end{subfigure}%
\begin{subfigure}{.5\textwidth}
  \centering
  \includegraphics[width=0.65\linewidth]{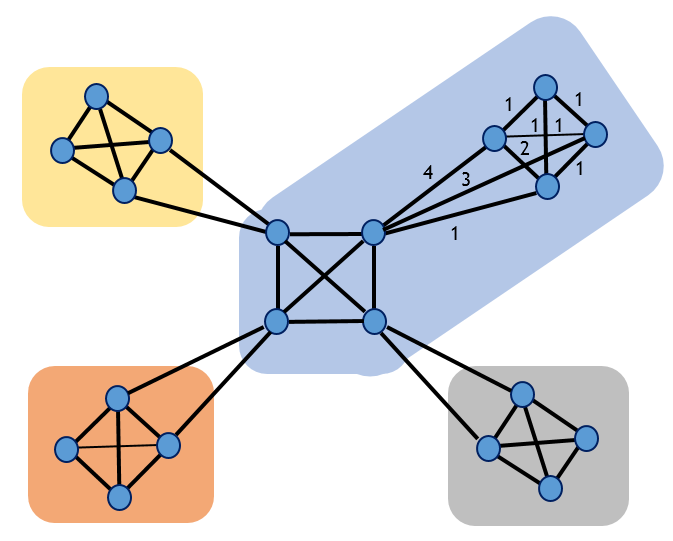}
  \caption{After adding the green edges}
  \label{fig:AAA}
\end{subfigure}
\caption{Edge addition scenario example}
\label{fig:AA}
\end{figure}

For this case, if \textit{newEdge} is inserted inside a community, this will not affect the community partition in terms of weights. Unlike, an inter-community edge could increase the inter-community weight, noted $ c_{1} $ and $ c_{2} $, aggregating them in one community. To handle this case, \textit{Dyci} verifies whether Equation \ref{merginCondition2} holds or not. The edge addition update scenario is illustrated with an example in Figure \ref{fig:AA}. 

\begin{equation}
\label{merginCondition2}
INW_{c_{1}, c_{2}} \geqslant IW_{c_{1}} \mbox{ or } INW_{c_{1}, c_{2}} \geqslant IW_{c_{2}}
\end{equation} 

\subsubsection{ EdgeWeightUpdating (\textit{edgeWeightUpdate}):}
\begin{figure}[!h]
\centering
\begin{subfigure}{.5\textwidth}
  \centering
  \includegraphics[width=0.65\linewidth]{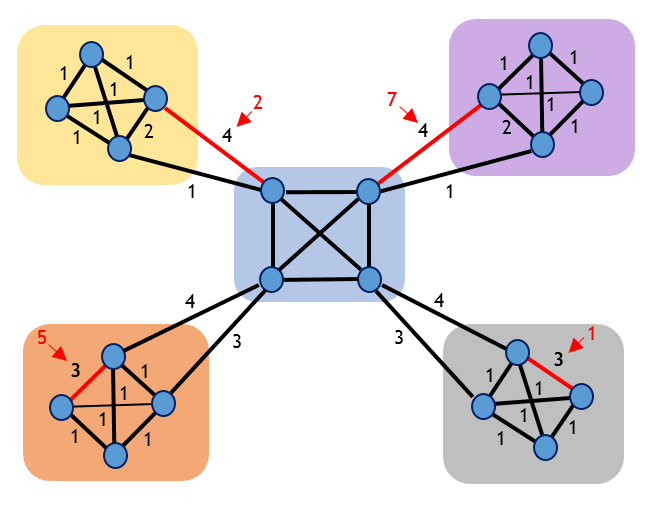}
  \caption{Before updating the red edges weights}
  \label{fig:WAB}
\end{subfigure}%
\begin{subfigure}{.5\textwidth}
  \centering
  \includegraphics[width=0.65\linewidth]{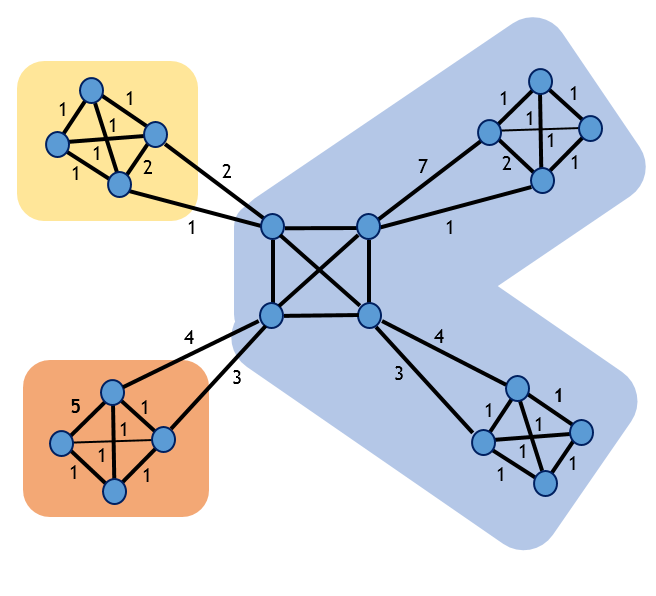}
  \caption{After updating the red edges weights}
  \label{fig:WAA}
\end{subfigure}
\caption{Edge weight update scenario example}
\label{fig:WA}
\end{figure}

The last and not least update case can be partitioned into two subcases, illustrated in Figure \ref{fig:WA}. The first subcase is when the \textit{edgeWeightUpdate} is an inter-community edge with weight greater than the old edge weight. For this scenario \textit{Dyci} verifies whether Equation \ref{merginCondition2} holds or not. The second subcase rises when \textit{edgeWeightUpdate} is an intra-community edge with weight lower than the old edge weight. The algorithm checks whether this weight loss leads to an adjacent community merging by Equation \ref{merginCondition1}.

\section{Genetic algorithm}
\label{GA}
Genetic algorithms (GA) can provide very good results if they are well set. In order to evaluate the quality of the obtained communities of $ Cs_{f} $, a comparison is conducted between \textit{Dyci} and the following GA.

\begin{figure}[!b]
\begin{center}
\includegraphics[scale=0.28]{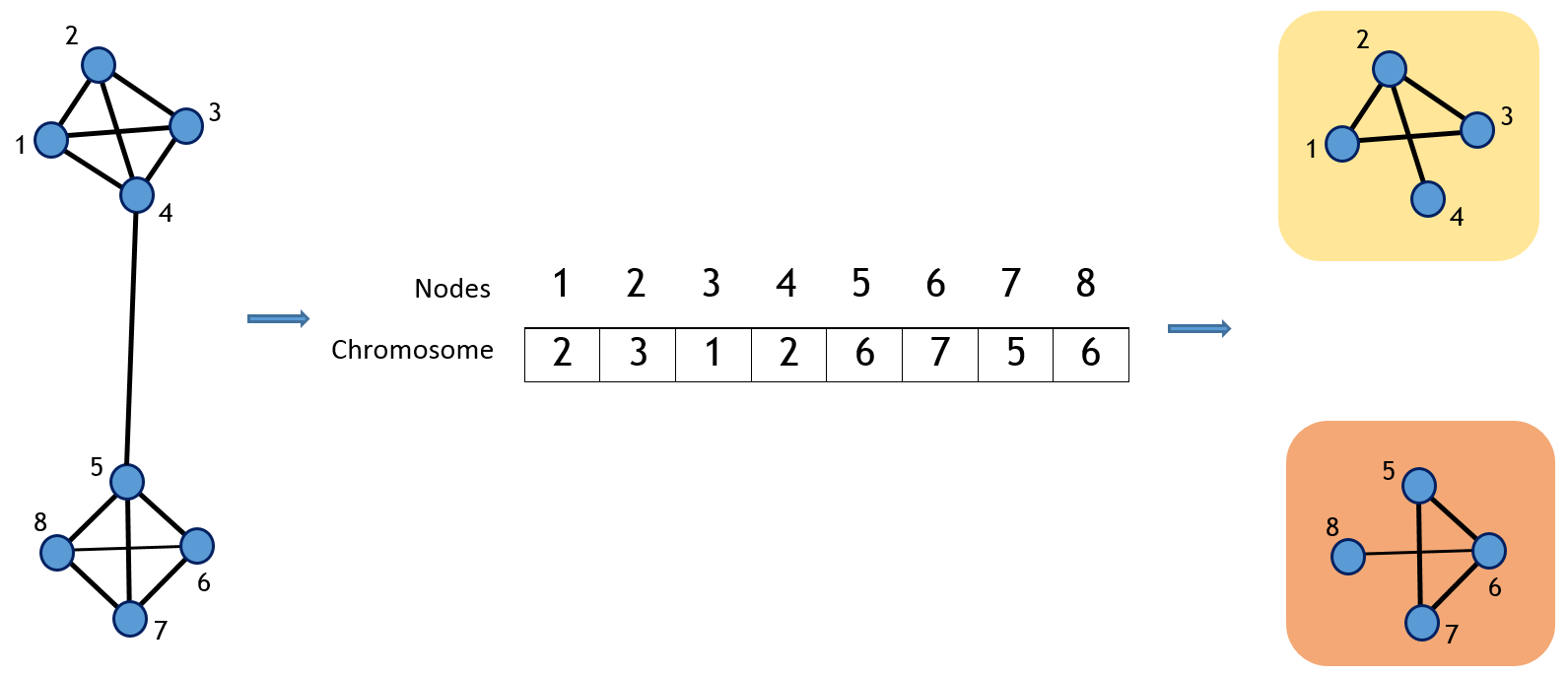}
\end{center}
\caption{An individual example using LAR encoding}%
\label{lar}%
\end{figure}

\begin{figure}[!t]
\centering
\begin{subfigure}{.5\textwidth}
  \centering
  \includegraphics[width=1.0\linewidth]{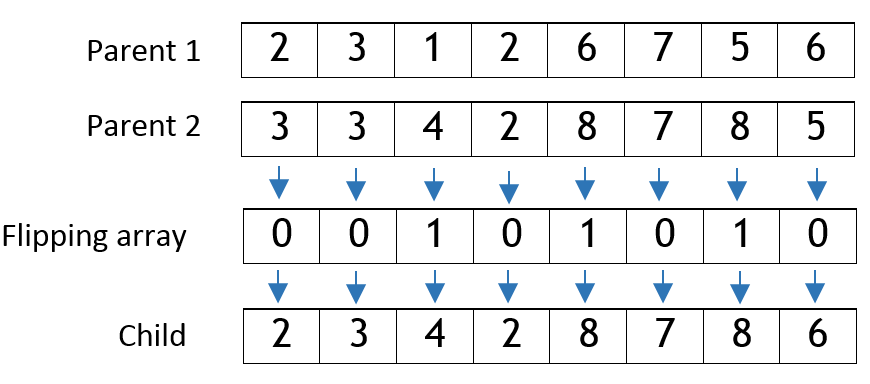}
  \caption{Uniform crossover}
  \label{fig:subCros}
\end{subfigure}%
\begin{subfigure}{.5\textwidth}
  \centering
  \includegraphics[width=1.0\linewidth]{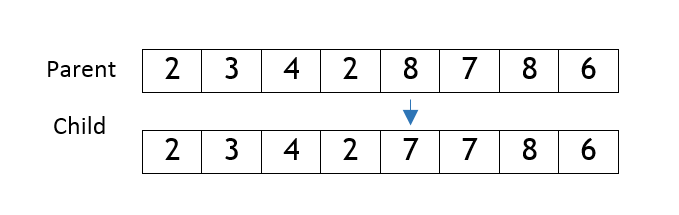}
  \caption{Mutation}
  \label{fig:subMut}
\end{subfigure}
\caption{Reproduction operators}
\label{fig:crosMus}
\end{figure}

\begin{itemize}
\item \textbf{Chromosome encoding}:
The Locus-based Adjacency Representation \cite{park1998} (LAR) is used to encode the community detection problem, like in \cite{PizzutigaNet}, \cite{JinHLB10}. In the LAR a $ |N_{f}| $ sized array is used, where the couples (gene, allele) express an associative community membership. Indeed, each gene takes its allele value from the set of its node neighbours ensuring feasible solutions. Figure \ref{lar} shows an example with the related individual decoding.

\item \textbf{Fitness function}:
Modularity $ \varphi $ of \cite{newman2006} defined by Equation \ref{eq:mod1} is used for individual evaluation.
\begin{equation} \label{eq:mod1}
 \varphi = \frac{1}{2M} \sum^{v}_{i=1} \sum^{v}_{j=1} \left( e^{w}_{n_{i}, n_{j}} - \frac{WD_{n_{i}} WD_{n_{j}}}{2M}  \right) \delta\left( c_{n_{i}}, c_{n_{j}}\right)
\end{equation}
Where:
\begin{itemize}
\item{} $  M = \sum^{v}_{i = 1} \sum^{v}_{j = i+1} e^{w}_{n_{i}, n_{j}} $
\item{} $ \delta\left( c_{n_{i}}, c_{n_{j}}\right) = 1, $ if $  c_{n_{i}} = c_{n_{j}}, 0  $ otherwise.
\end{itemize}
The modularity expresses whether the detected community structure is well defined or not, corresponding to the density of the detected communities minus the density of these communities for the random case with the same characteristics. As an instance, let suppose that the edge weight set for the graph of Figure \ref{lar} is defined as follows: 
\begin{itemize}
\item{} $  \lbrace E^{w}_{t} \rbrace^{|E_{t}|} = 1 $, thus $  M = 13 $ and $ \varphi = 0.423.$
\end{itemize}
\item \textbf{Population initialization}: 
A random population of size $ 100 $ is generated and sorted in a decreasing fitness function order.
\item \textbf{Crossover}: 
Uniform crossover with probability $ 0.9 $ is performed, as illustrated in Figure \ref{fig:subCros}.
\item \textbf{Mutation}:
Random allele flipping with probability $ 0.1 $ is performed, as showed in Figure \ref{fig:subMut}.
\item \textbf{Parent selection and child insertion}: 
Random selection from the $ 20\% $ eliteness individuals. Weakest individuals are excluded from the population.
\item \textbf{Stopping condition}:
Number of generations reaches $ 50 $.
\end{itemize}

As previously mentioned, the LAR is used as encoding because it ensures feasible solution solutions with the induced (gene, allele) associative community membership and because it provides a straightforward decoding step (i.e. linear with respect the size of the individual chromosome). Additionally, the chosen crossover and mutation operators maintain the feasibility solution of the LAR encoding.  

\section{Twitter's network and experimentation}
\label{experiments}

\begin{figure}[!t]
\centering
\begin{subfigure}{.9\textwidth}
  \centering
  \includegraphics[width=0.85\linewidth]{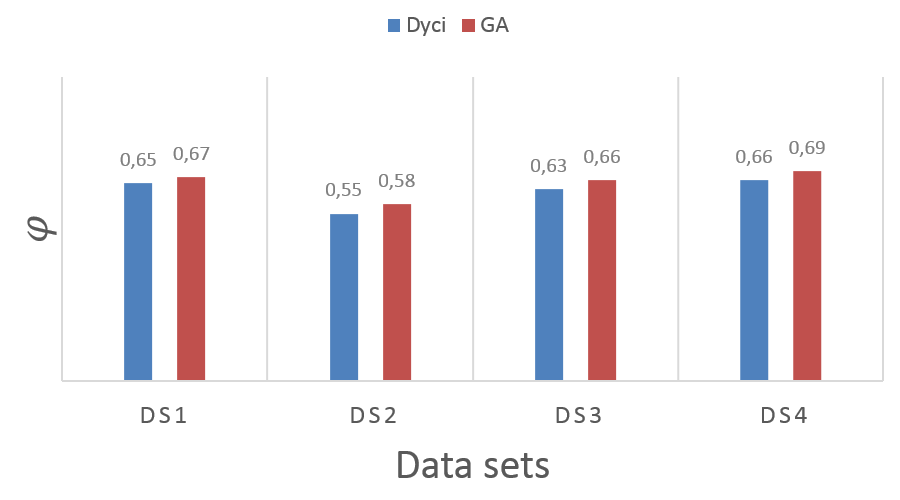}
  \caption{Obtained modularity}
  \label{fig:modDynamic}
\end{subfigure}%
\newline
\begin{subfigure}{.9\textwidth}
  \centering
  \includegraphics[width=0.85\linewidth]{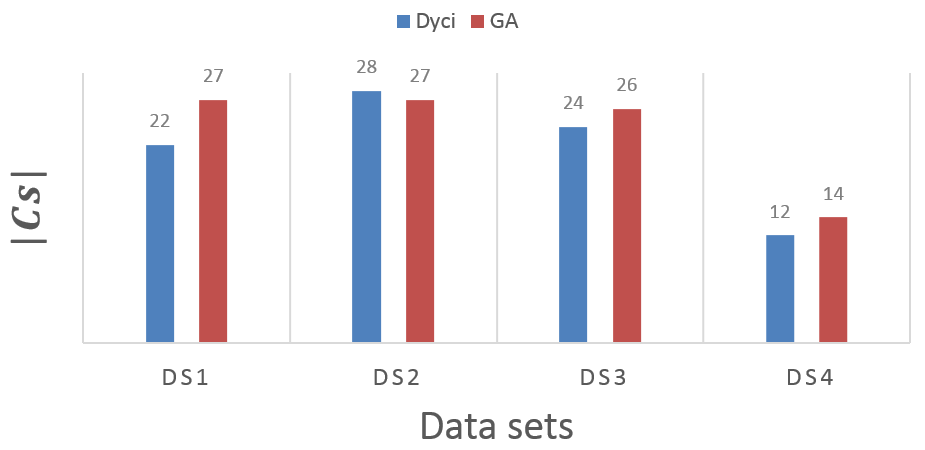}
  \caption{Identified number of communities}
  \label{fig:csDynamic}
\end{subfigure}
\begin{subfigure}{.9\textwidth}
  \centering
  \includegraphics[width=0.85\linewidth]{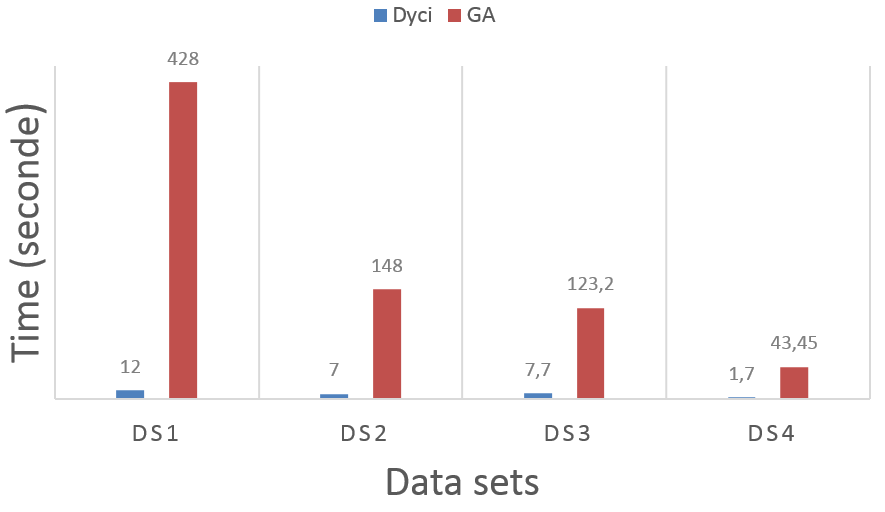}
  \caption{Running time}
  \label{fig:timeDynamic}
\end{subfigure}
\caption{The results for the ANR-Info-RSN data sets at $ t_{f} $}
\label{fig:lastDynamic}
\end{figure}

\begin{figure}[!t]
\centering
\begin{subfigure}{.9\textwidth}
  \centering
  \includegraphics[width=0.82\linewidth]{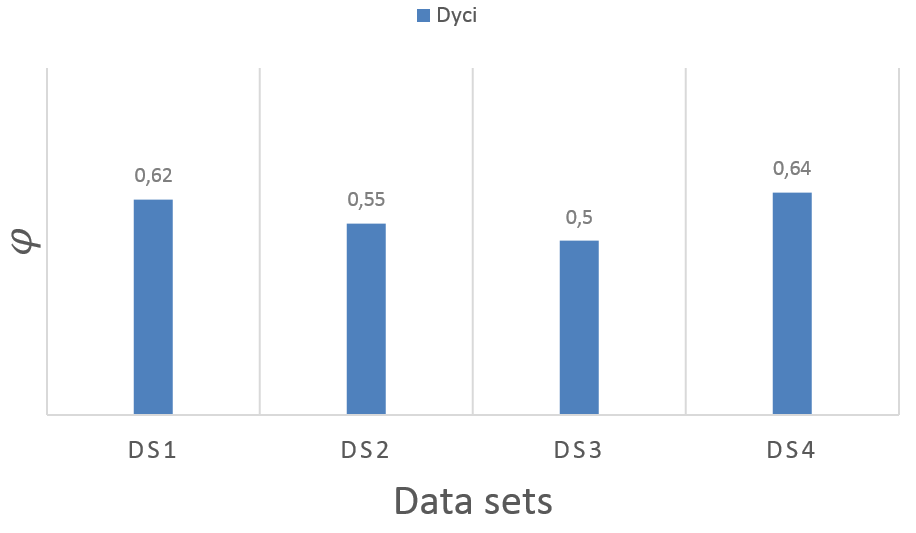}
  \caption{Obtained modularity}
  \label{fig:dyciMod}
\end{subfigure}%
\newline
\begin{subfigure}{.9\textwidth}
  \centering
  \includegraphics[width=0.82\linewidth]{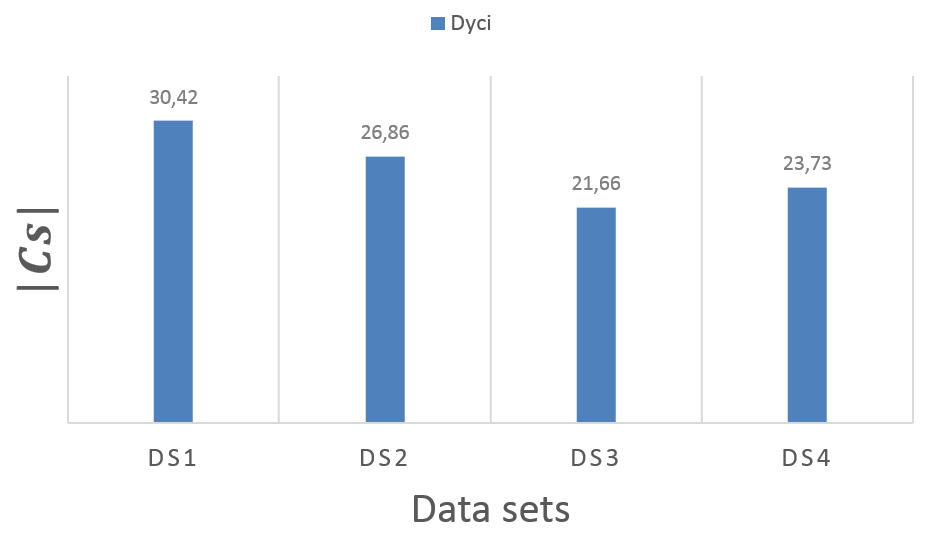}
  \caption{Identified number of communities}
  \label{fig:dyciCom}
\end{subfigure}
\begin{subfigure}{.9\textwidth}
  \centering
  \includegraphics[width=0.82\linewidth]{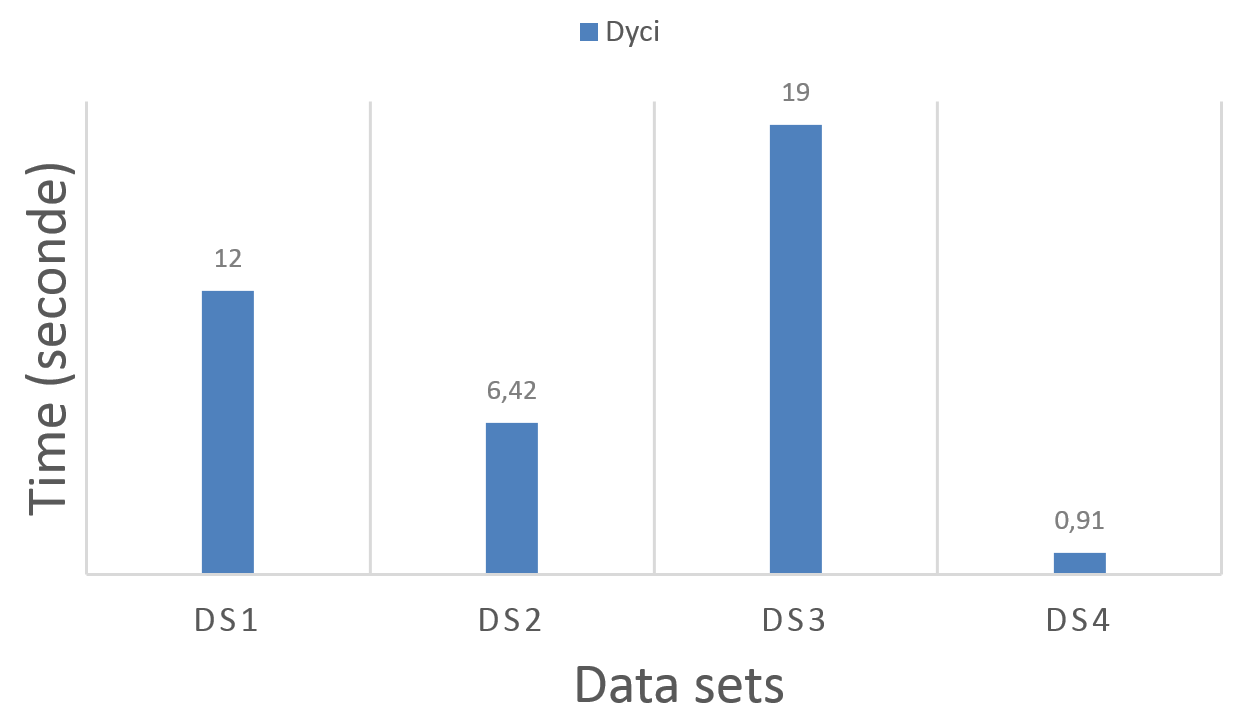}
  \caption{Running time}
  \label{fig:dyciTime}
\end{subfigure}
\caption{The results for the ANR-Info-RSN data sets for the whole $ Cs_{s} $}
\label{fig:dyciWhole}
\end{figure}

This section discusses the obtained results of the conducted comparison between the previous GA and \textit{Dyci} on four datasets from real-world data of the ANR-Info-RSN project. The ANR-Info-RSN project deals with the community detection in Twitter, especially in a huge collected set of tweets from social media. To this end, a graph is used as model, where each Twitter's user of the collected data is represented by a node and an edge represents a retweet relationship between two Twitter's users. In this context, the edge weight is equal to the number of times where a retweet is observed between two Twitter's users. Table \ref{tab:dynamicChar} presents the datasets characteristics where the unit of snapshot generation is one day. Figure \ref{fig:lastDynamic} and Figure \ref{fig:dyciWhole} show, respectively, the obtained results for the datasets at $ t_{f} $ and their averages values for the datasets for the whole $ Cs_{s} $.

\begin{table}[!t]
\caption{The ANR-Info-RSN datasets characteristics} 
\renewcommand{\arraystretch}{1.2}
\centering\begin{tabular}{ccccccc}
\hline\noalign{\smallskip}
Data sets & $ t_{0} $ & $ t_{f} $ & $ |N_{s}| $ & $ |E_{s}| $ & $ |N_{f}| $ & $ |E_{f}| $ \\
\noalign{\smallskip}
\hline
\noalign{\smallskip}
DS1 & July 17,2014 & July 31,2014 & 10569 & 14121 & 801 & 997 \\
DS2 & August 3,2014 & August 15,2014 & 6162 & 8069 & 390 & 451 \\ 
DS3 & August 17,2014 & August 31,2014 & 10189 & 12263 & 424 & 508 \\
DS4 & September 3,2014 & September 30,2014 & 8224 & 10371 & 412 & 535 \\
\hline
\end{tabular}
\label{tab:dynamicChar}%
\end{table}

From Figure \ref{fig:modDynamic}, we remark that \textit{Dyci} and the GA have almost the same results (GA very slightly better), taking into account the fact that the obtained communities $ \mathit{Cs_{f}} $ of \textit{Dyci} are highly influenced by the $ f $ previous choices made during the whole graph sequence. One could say that \textit{Dyci} obtains satisfactory results. Further, from Figure \ref{fig:timeDynamic}, we remark that \textit{Dyci} is relatively fast compared to the GA, due to the fact that \textit{Dyci} takes advantage from the previous identified community avoiding relaunching the process at each snapshot. From Figure \ref{fig:dyciWhole}, we notice that the averages values are almost the same by comparing to the values of the last snapshot $ t_{f} $, except for DS3 where \textit{Dyci} takes more time and provides less modularity for the previous snapshots but has relatively a good result for the last snapshots $ t_{f} $.

\section{Dynamic community visualization with \textit{NLCOMS}}
\label{vis}

After handling the set of updates $ \mathit{U_{t}} $ and thus, inferring the impacts that cause to the community set leading to $ \mathit{Cs_{t+1}} $, one has to draw the related graph $ \mathit{G_{t+1}} $. In this work, we use the previous layout $ \mathit{L_{t}} $ of  $ \mathit{G_{t}} $ as a starting point to embed the graph  $ \mathit{G_{t+1}} $ with $ \mathit{L_{t+1}} $. To this end, we use our visualization tool, called \textit{NLCOMS} (Node-Link and COMmunitieS), which relies on node-link diagram and force-directed algorithm \cite{Kobourov13} for dynamic graph drawing while providing three node positioning possibilities: 

\begin{itemize}
\item \textbf{Free}: the charges of the force-directed algorithm are applied on all the nodes which induces a freely node positioning.  
\item \textbf{Fixed}: The nodes present in the previous layout $ \mathit{L_{t}} $ are kept in the same position in $ \mathit{L_{t+1}} $. 
\item \textbf{Anchored}: This idea was already introduced by \cite{BrandesW97} in their absolute paradigm. Here, a virtual node is linked to each node which was in the previous layout $ \mathit{L_{t}} $ acting like a boat anchor.
\end{itemize}

\begin{figure}[!b]
\begin{center}
\includegraphics[scale=0.85]{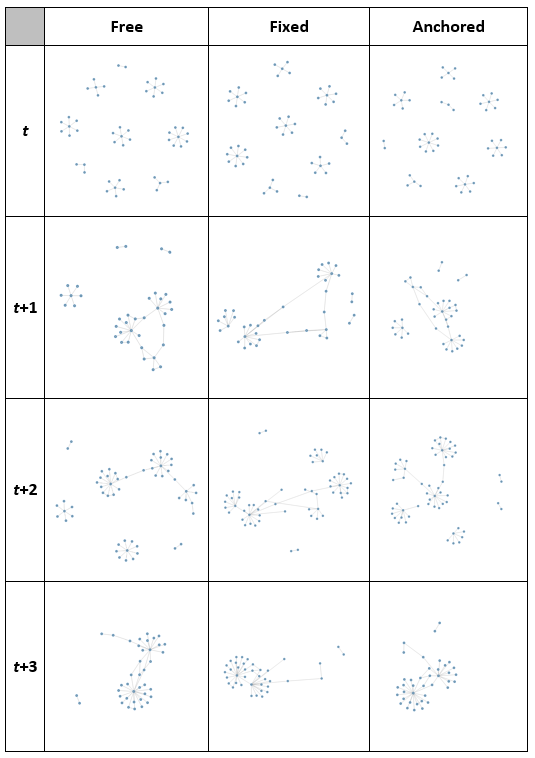}
\end{center}
\caption{A comparison between node position possibilities drawn using \textit{NLCOMS}}%
\label{anchComp}%
\end{figure}

Figure \ref{anchComp} illustrates a sample graph comparison of the previous three node positioning possibilities using \textit{NLCOMS}. From Figure \ref{anchComp} we can remark that the three node positioning possibilities behave differently. If the user's visual task is to detect a specific node through the different snapshots, it is obvious that the freest way to layout node might lead to misleading interpretation with temporal alias \cite{beck2009} (i.e., in the node-link context, nodes mistaken one for the other  due to their positioning in different snapshots). The fixed node position is efficient for relatively small graphs, however it suffers from scalability issue. Additionally, the final node position strongly depends on the initial node position. As an example, in the layout at $ \mathit{L_{t+2}} $, the uniformity of the edge length and the edge crossing criteria are not met. The third possibility, the anchored node position, provides a pleasing graph drawing manner and ensures a good trade-off between user's mental map preservation and graph drawing aesthetics.

\begin{figure}[!t]
\begin{center}
\includegraphics[scale=0.35]{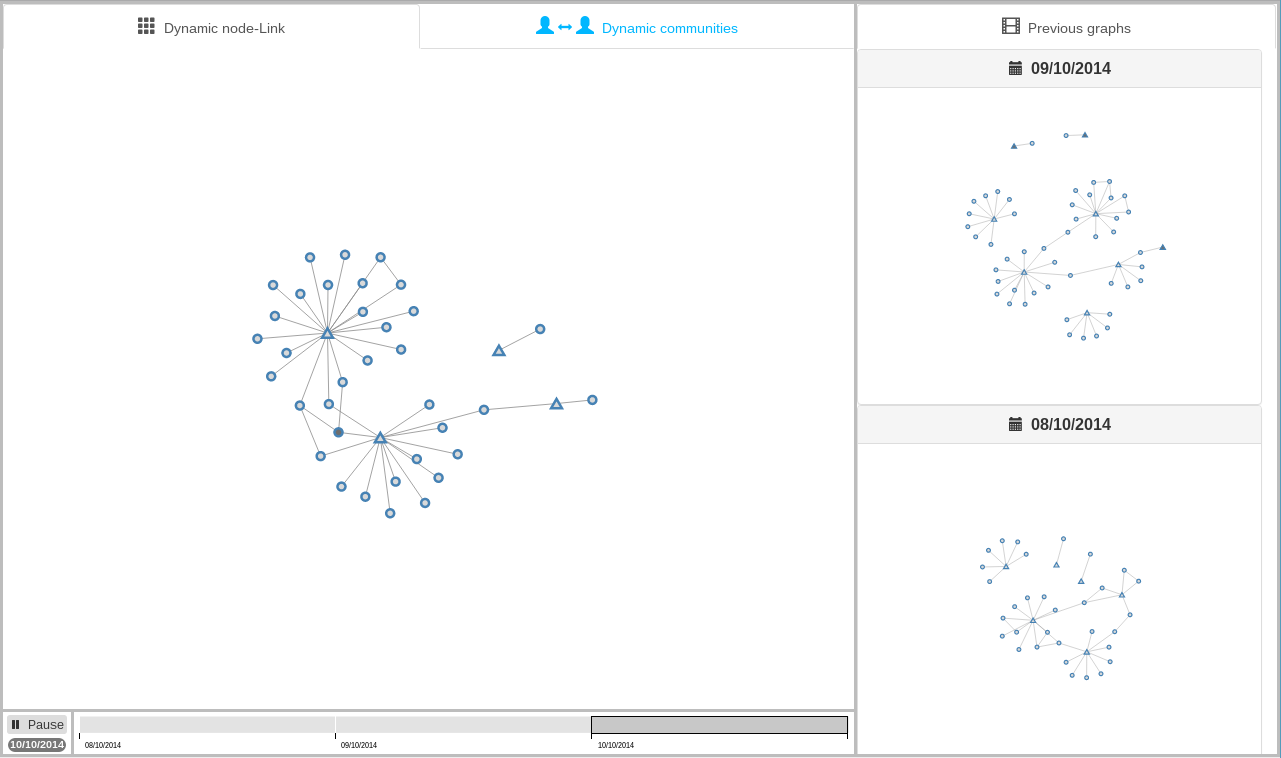}
\end{center}
\caption{Multiple views of \textit{NLCOMS} for dynamic graph visualization}%
\label{nlcomsDynamic}%
\end{figure}

The next step of the proposed approach in this work consists in displaying the successive graph layouts in the limited depiction area. To this end, \textit{NLCOMS} uses the physical time and the axial time in order to take advantage from both of them. Figure \ref{nlcomsDynamic} shows a sample from the real-world data of ANR Info-RSN database from October 8, 2014 to October 10, 2014. The two last snapshots are depicted allowing the user to scroll down and up to explore the historic of the graph structure. As the graphs are generated from the initial graph $ \mathit{G_{0}} $, the user would like to perceive the presence of the initial nodes. For that, the shape is used as a visual variable \cite{Bertin1967} to fulfill this task, represented with triangles. Additionally, \textit{NLCOMS} highlights nodes' presence frequency over time via the node inner color saturation. Indeed, the darker a node is, the more frequent it was observed through the previous snapshots. This visual task provides to the user a perception of nodes persisting over successive snapshots and those which are just outliers at a specific snapshot. However, even if they are outliers at a specific instant $ \mathit{t} $, they may contribute to the overall graph structure understanding. Finally, a different color is assigned to the nodes which belong to different communities detected by \textit{Dyci}. This ensures community membership distinction as user visual task. 

\section{Conclusion}
\label{conlusion}

Dynamic social network analysis attracts more interest from the scientific community. Indeed, understanding of the graph structure over time is a very challenging task. In this work, we propose an approach for analysing dynamic social networks, more precisely for Twitter's network. Our approach relies on two complementary steps: (i) an online community identification based on a dynamic community detection algorithm called \textit{Dyci}. The latter checks over time whether a connected component of the weighted graph becomes weak in terms of weight, in order to proceed to a local community re-identification. Thus, community detection from scratch at each snapshot is avoided. (ii) a community visualization is provided by \textit{NLCOMS}, which combines two methods (i.e., physical and axial time) of dynamic network visualization. The efficiency of the proposed approach is assessed and a comparison study is conducted with a genetic algorithm on samples from real-world data of the ARN-Info-RSN project, which deals with community analysis in Twitter. The results show that our proposed approach allows us to efficiently detect and to visualize the underlying communities to the network structure. 

As perspectives, we project to extend this work to multi-graphs with several edges types linking two nodes. As an instance, considering simultaneously retweet edges and "mention" edges. The latter exists when a Twitter's user quotes another Twitter's user in its tweet. In this context, community detection algorithm could consider overlapping communities. 

\section*{Acknowledgements}

This research has been supported by the Agence Nationale de la Recherche (ANR, France) during the Info-RSN Project (ANR-13-SOIN-0008).

\section*{References}
\bibliography{mybibfile}

\end{document}